\documentclass[preprint]{aastex}

\newcommand{\boo}{Bo\"{o}tes I}

\begin{document}





\title{The Age and Metallicity of the  Bo\"{o}tes I System}


\author{Joanne Hughes}
\affil{Physics Department, Seattle University, Seattle, WA 98122}
\email{jhughes@seattleu.edu}
\author{George Wallerstein}
\affil{Astronomy Department, University of Washington, Box 351580, Seattle,
WA 98195-1580}
\and 
\author{Anne Bossi}
\affil{Physics Department, Seattle University, Seattle, WA 98122}



\begin{abstract}
	We present Washington $CT_1T_2$ photometry of a field central to the \boo\ dwarf spheroidal
galaxy, which was discovered as a stellar overdensity in the Sloan Digital Sky Survey (DR5). We show that the Washington filters are much more effective than the Sloan filters in separating the metal-poor turn-off stars in the dwarf galaxy from 
the foreground stars. We detect 165 objects in the field, and statistically determine that just over 40\% of the objects are
non-members. Our statistical analysis mostly agrees with radial velocity measurements of the brighter stars.
We find that that there is a distinct main-sequence turn-off and subgiant branch, where there is some evidence of a spread in chemical abundance. Any evidence of an age spread is limited to a few billion years. The brightest 7 \boo\ members
give a (photometric-color derived) weighted mean iron-abundance of  $[Fe/H]=-2.1^{+0.3}_{-0.5}$, and the best-fit isochrone 
is the  14.1~Gyr, Z=0.0002 model, with $(m-M)_V=19.11$ and $E(B-V)=0.02$. 
\end{abstract}

\section{Introduction}

Over the last few years, the  Sloan Digital Sky Survey
(SDSS, York et al. 2000) has been mined for stellar overdensities, leading to the discovery of many new systems which have
the broad characteristics of dwarf galaxy-satellites of the Milky Way. In addition to \boo\ (Belokurov et al. 2006a), we also include Bo\"{o}tes II (Walsh et al. 2007), Canes Venatici (Zucker et al. 2006a), Willman 1 (Willman et al. 2005a), Ursa Major (Willman et al. 2005b), Ursa Major II (Zucker et al. 2006b), Hercules, Coma Berenices, SEGUE 1, Canes Venatici II, Leo IV (Belokurov et al. 2006b).  

Belokurov et al. (2006a, hereafter B06) describe the method used to 
search the SDSS $ugriz$ data for stellar overdensities. They also performed follow-up observations with the 4-m Blanco Telescope 
at the Cerro Tololo Inter-American Observatory in Chile of a $36^\prime \times 36^\prime$ field in the \it g \rm and \it i \rm bands. B06
compared the color-magnitude diagram (CMD) of \boo\ with that of the metal-poor galactic globular cluster, M92. Their deep (wide-field) \boo\ CMD is very crowded near the main sequence turn-off,
but B06 use the fiducial ridgeline of M92 to state that \boo\ is younger and slightly more metal-poor than M92 (for which  $[Fe/H]\sim -2.3$). The characteristic scale length of \boo\ was found to be about 220~pc at a distance of about 60~kpc. This distance is similar to that of the outermost halo globular clusters in our galaxy, closer than that of most of our accompanying dwarf galaxies which reside at distances of about 100~kpc (accompanied by a few globulars).
This preliminary work shows that the dwarf galaxy is
distorted (from the density contours), which suggests that it may be experiencing tidal disruption. At $M_V ~ -5.8\; mag$, \boo\ is one of the faintest
dSphs found to date, and one of the closest.

Mu\~{n}oz et al. (2006) took spectra of red giant branch (RGB) and asymptotic giant branch (AGB) stars (selected from SDSS DR4, because of
proprietary issues at the time) in the core of \boo\ and south of  declination $14.8^\circ$. The radial velocity data on the stars did not clearly 
distinguish the \boo\ population from the Milky Way stars, so they used spectroscopic features to remove foreground stars. Their data (for only 7 stars)
yielded a systematic velocity of $95.6\pm 3.4\; km/s$, a central velocity dispersion of $6.6\pm 2.3\;  km/s$, and a mass of $1.1^{+1.3}_{-0.5}
\times 10^7 M_\odot$. Mu\~{n}oz et al. (2006) have found that \boo\ is not only one of the faintest known dSphs, but also one of the darkest
(M/L ratio of between 130--680),
and most metal poor, at $[Fe/H]\sim -2.5$. Martin et al. (2007) observed candidate \boo\  red giants from SDSS (DR4) with Keck/DEIMOS, converting
measurements of the Ca~II lines to $[Fe/H]$, finding $99.9\pm 2.4\; km/s$, with central velocity dispersion $\sigma = 6.5^{+2.1}_{-1.3}\; km/s$ for their
final sample of 24 stars with small velocity uncertainties. For the DEIMOS sample, $[Fe/H]\sim -2.1$, with one star at -2.7. Martin et al. (2007) discuss the
systematic uncertainties involved, and part of the reason their data seems to skew more metal-rich than Mu\~{n}oz et al. (2006) (and Siegel 2006) 
may be due to Martin et al. using the Caretta \& Gratton (1997) metallicity scale rather than that of  Zinn \& West (1984). However, Martin et al. note
that none of the groups doing spectroscopy were using techniques which had been calibrated below $[Fe/H]\sim -2.3$, so the discrepancies might
stem from that factor.

Siegel (2006) and Dall'Ora et al. (2006) have studied variable stars in \boo . Using the McDonald Observatory 0.8-m telescope and B, Washington M, $T_2$ (I-band), and DDO51 filters, Siegel (2006) identified 15 RR Lyrae stars, and estimated the
distance to \boo\ as $ (m-M)_0=18.96 \pm 0.12$, using $E(B-V)=0.056$, with  $-2.5< [Fe/H]< -2.0$. From 58 B-images, Siegel found the average $RRab$ period to be 0.69 days (period shift -0.07, compared with M3 RR Lyraes), with 53\% being type c pulsators, and classifying it as an Oosterhoof type II system (rather than \it intermediate, \rm as for other dSphs, except Ursa Minor), which also implies it is very metal-poor. Siegel's CMD was in (B-I), with the M-DDO51 colors used to remove field dwarfs. Dall'Ora et al. (2006) presented the V,I light curves of 12 variables, 11 of which were RR Lyraes. They find the distance modulus to be $19.11\pm 0.08\; mag$ for the assumed metal abundance of  $[Fe/H]\sim -2.5$ and $E(B-V)=0.02$, which we use
as our distance modulus and extinction values in this paper. Unfortunately, these variable stars fell outside our central field.

Recently, de Jong et al. (2008) performed a numerical analysis of the SDSS CMDs of several dSphs and globular clusters, using the new software package MATCH. They found that the \boo\  SDSS data was consistent with the population being old and metal-poor, with no evidence of more than one burst of star formation. The CMD-analysis only showed CVn I, UMa II, and Leo T having more than one epoch of star formation, but the conclusions were drawn from SDSS-data only,
 
 We have obtained photometry in the Washington filters to investigate the main sequence turn-off (MSTO) and subgiant branch (SGB) of \boo . The $CT_1T_2$ photometry provides metallicity as well as temperature information for the cluster's giants and subgiants. The position of the RGB is metallicity-dependent, but  we examine the MSTO/SGB region  to search for evidence of age spreads.

\section{Observations}

	We observed one field centered on ($RA=14^h 00^m 06^s$, $Dec= 14.5^\circ$; \it J2000)\rm\  with the  Apache Point Observatory 3.5-m telescope, using   the direct imaging SPIcam system. The detector is a backside-illuminated SITe TK2048E $2048\times 2048$ pixel CCD with 24 micron pixels, which we binned ($2\times 2$), giving a plate scale of 0.28 arcseconds per pixel, and a field of view of $4.78\times 4.78$ square arcminutes. The data set for  \boo\ was taken on 2007 March 19. We took 21 frames in Washington C, and Cousins R 
and I filters, with exposure time ranging from 1 seconds to 1000 seconds. The readout noise 
was 5.7e- with a gain of 3.4 e-/ADU.  The images were 
flat-fielded using dome flats, along with  a sequence of zeros. We then 
processed the images using the image-processing software in IRAF.

As we described in Hughes et al. (2007; studying the bulge globular cluster, NGC 6388), we substituted R- and I-filters for Washington $T_1$ and $T_2$, because $(R-I)$ can be converted to $(T_1-T_2)$ 
linearly. R \& I are broader than Washington $T_1$ and $T_2$, which reduces the 
observing time, and the Washington C-filter is much broader than the B-band of the UBV system as well
as the Str\"{o}mgren v-band.  The C-band is also more sensitive to metallicity than the B-filter. 
Due to these factors, we prefer the Washington filters over the more commonly used BVRI-filters. 
Geisler \&  Sarajedini (1999) explained the
advantages that the Washington system provides:
they widen the separation of the giant branches of different metallicities (giving a resolution for
RGB fiducials of $\sim 0.15\; dex$), while the reddening 
sensitivity of the Washington filters is half that of the (V-I) colors. In this paper, we use  the ``New Washington System CCD Standard Fields", SA 101 and SA 107 (Geisler 1996),
the standard giant branches of Geisler \&  Sarajedini (1999; hereafter, GS99), 
and the theoretical models of Marigo et al. (2008).  We note that Sneden et al.  (2000) find that M15 and M92 have a  similar abundances, $[Fe/H]\sim -2.3$, wheras the -2.15 of GS99 comes from Zinn (1985).

Our photometry on 2007 March 19 yielded matches to the GS99 standard system of
$\sigma_{rms}=0.011$ in $T_1$, $\sigma_{rms}=0.023$ in $C$, and $\sigma_{rms}=0.017$ in $T_2$. 
In $T_1$, the average uncertainties in the final CMD were $\sigma_{rms}\leq 0.03$ at the level of the horizontal branch, 
and $\sigma_{rms}=0.05$ just above the MSTO. The transformation equations are as follows:
\begin{equation}
T_1=R_i-0.289+0.023(C_i-R_i)-0.206X, \; \sigma_{rms}=0.011
\end{equation}
\begin{equation}
C=C_i-1.320+0.128(C_i-R_i)-0.396X, \; \sigma_{rms}=0.023
\end{equation}
\begin{equation}
T_2=I_i-0.768+0.006(R_i-I_i)-0.087X, \; \sigma_{rms}=0.017
\end{equation}
Here, \it X \rm denotes the airmass and the subscript \it i \rm indicates the instrumental magnitude.

Table 1 lists the data taken in spring, 2007 at APO.
The images taken on 2007 March 19 had sub-arcsecond seeing. We did take data on an offset \boo\ field and an off-dSph region at the same galactic latitude on the night of UT 2008 January 14, but the seeing was 2-3$^{\prime\prime}$ and variable. We use the ``foreground'' field for non-member selection, but we do not include the extra \boo\ field because the data were of much poorer quality.
 
 We used the DAOPHOT 
program in IRAF (Stetson et al. 1990) to perform crowded field photometry, although it was scarcely necessary for this
sparsely populated field (compared to a globular cluster) of the \boo\ dwarf. Artificial-star experiments found that for the deepest
C-frames, and the 300-s R and I exposures, nearly all the artificial objects placed in the field (except close to the bright foreground stars) were  recovered (adding about 20\% to the total number
of detections, with a range of magnitudes from 16--22 in $T_1$).

We used two iterations of (DAOPHOT-PHOT-ALLSTAR), with the first iteration having 
a detection threshold of 4$\sigma$, and the second pass had a 5$\sigma$ detection limit. We used 10-15 stars to 
construct the point spread functions (PSFs), and did not allow it to vary over the chip (there was clear evidence that the PSF was constant over the whole frame anyway).
ALLSTAR was further constrained to only detect objects with a  CHI-value 
(the DAOPHOT goodness-of-fit statistic) between 0.5 
and 1.5 (to remove cosmic rays and non-stellar, extended objects). 
We found the aperture correction between the small (3 pixel) aperture used 
by ALLSTAR in the \boo\ field, and the larger (10 pixel) aperture used for the 
standards, by using the PSF stars in each image. 
We used the IMMATCH programs
 to match the source lists in each frame, which resulted in several data sets in C, $T_1$ \&  $T_2$. We then put together the final 
source list as follows: requiring that each star be detected in at least one image in each filter, and  the final magnitude and colors were calculated as the weighted mean of each individual 
detection
We calculated uncertainties for each individual object in 
each frame by taking the uncertainties from photon statistics, DAOPHOT's uncertainties, the 
aperture corrections, and the standard photometric errors in quadrature. Figure~1 shows the total
uncertainties in $T_1$, $(C-T_1)$ and $(T_1-T_2)$ against the  $T_1$-magnitude.

\section{\boo\  Membership}

Figure 2a shows  Washington CMD of the 165 stars (listed in Table 2). In Figure 2b, we display the data set with the standard giant branches
of GS99 (having marked values of [Fe/H]), where M15 (GS99) data (small circles) are shown for comparison.
We use the distance modulus of 19.11 and $E(B-V)=0.02$. To convert from $E(B-V)$ to the reddening in the Washington filters we used:
$E(C-T_1) = 1.966E(B-V)$;
$E(T_1-T_2) = 0.692E(B-V)$;
$M_{T_1} =T_1+ 0.58E(B-V) - (m-M)_V$;
from Geisler, Claria \& Minniti  (1991) and GS99. We note that care should be taken in areas of 
suspected variable extinction (see discussion in Hughes et al. 2007, for NGC 6388). Twarog et al. (2006) state that a small shift of only 0.05 in $E(B-V)$ increases the Washington-$[Fe/H]$ from $-0.57$ to $-0.35$ for Melotte 71.  Fortunately, the small extinction value of $E(B-V)\sim 0.02$, (Dall'Ora et al. 2006) and high galactic latitude, reduces this possibility for \boo . We plot the \boo\ stars (open squares) with their error bars, and note that there are at least 4-6 stars in the blue straggler region, brighter and bluer than the main-sequence turn-off. Our photometry does not appear to have any color-shifts with respect to the M15 data from GS99,  which has $(m-M)_V=15.41$ and $E(B-V)=0.10$.

Our method of using three Washington filters, $CT_1T_2$, is meant to provide more than one color to determine  \boo\ membership. As demonstrated by Siegel, Shetrone \& Irwin (2008) for Willman~1, follow-up studies  showed that many of the brightest stars thought to be part of that system's RGB, were dwarf stars and halo objects. For future follow-up spectroscopy, we want to know whether the \boo\ RGB stars really belong to the dSph. 

In an ideal situation, we would have obtained exactly the same number of frames in each filter on the off-galaxy field (at the same galactic latitude), on the same night, under the same seeing conditions.  However, we  obtained  frames of an offset \boo\ field and an off-galaxy field on 2008 January 14, with integration times of 180s at I, 240s at R and 360s at C. These data only reached the base of the RGB, with seeing of 2--3 arcseconds, compared to sub-arcsecond seeing on 2007 March 19.  We did detect
14 objects in the off-galaxy field compared with 34 in the offset \boo\ field. We show these objects as 
open stars in Figure~3. This CMD also shows the M15 objects (GS99, small circles) and the proposed 
\boo\ stars (filled squares).  Since the off-galaxy field could not go deep enough to reach the \boo\
MSTO, we used the TRILEGAL 1.1 code (see Girardi et al. 2005), which simulates photometry (in many
filter combinations) for any galactic field. We ran the code, using the web interface,\footnote{http://stev.oapd.inaf.it/cgi-bin/trilegal}
for the same area as our \boo\ field, with the same limiting magnitude in $T_1$.  The run with the standard galactic parameters ($A_V$, IMF, e.t.c., see  Girardi et al. 2005) yielded 76 simulated objects
(shown as crosses in Figure~3), which can now be treated as foreground contaminating stars. From the
initial numbers in the real and simulated fields, we expect at least 40\% of the on-\boo\ source list to be foreground objects, mostly in the giant branches, not the MSTO region. As a first cut in the \it cleaning
\rm process, Figure~3 shows that the M15 data and the \boo\ MSTO are best fit by the Z=0.0002, 
14.1~Gyr isochrone of Marigo et al. (2008).\footnote{http://stev.oapd.inaf.it/cgi-bin/cmd} It is likely that none of the \boo\ members fall beyond the 
Z=0.0006, 12.7~Gyr isochrone.

Figure 4 is the $(T_1-T_2)$ vs.  $(C-T_1)$ color-color plot of the \boo\ stars with uncertainties better than 0.05, shown as filled squares, the off-galaxy field objects are open stars, and the simulated foreground objects are crosses. The constant-$[Fe/H]$ lines from Geisler, Claria \& Minniti  (1991) are marked.  The  Z=0.0006, 12.7 Gyr, isochrone of Marigo et al. (2008) is also drawn, showing that the sub-giant and 
MSTO stars that do not share the colors of the foreground objects, stay to  the metal-poor side of this line.
This color-color plot is only sensitive to metallicity, as all the isochrones at the same Z fall in the same locus.
Foreground dwarfs with red $(T_1-T_2)$ vs.  $(C-T_1)$ colors do occupy the locus of the metal-poor 
RGB-tip stars, which was the case with Willman~1 (Siegel, Shetrone \& Irwin 2008). Thus, Figure~4 
shows that resorting to cuts in color would only be successful at the MSTO to the base of the RGB.
As we have done with NGC 6388 \& $\omega$~Cen (Hughes et al. 2007; Hughes \& Wallerstein 2000),
we can statistically compare the off-galaxy region (and the simulated field) to the
\boo\ field.   We subtract the foreground stars statistically by the following method, used by Hughes \&  
Wallerstein (2000), which was adapted from Mighell, Sarajedini, \&  French (1998). 

We calculate the probability that the star in the \boo\ field CMD 
is a member  of the dSph population as: 
\begin{equation}
p\approx 1-min\left({\alpha N_{off}^{UL \; 84}\over N_{on}^{LL\; 95}},\; 1.0\right)
\end{equation}
			                  
Where $\alpha$ is the ratio of the area of the dSph galaxy region to the area of the field region and
\begin{equation}
N_{off}^{UL \; 84}\approx (N_{off}+1)\left[  1- {1\over 9(N_{off}+1)}
+{1.000\over 3 \sqrt{N_{off}+1}}\right ]^3
\end{equation}
The equations are taken from the Appendix of Hughes \&  Wallerstein (2000), and 
corresponding to eq. [2] of Mighell et al. (1998) and eq. [9] of Gehrels (1986). Here, Equation 5 is the 
estimated upper (84\%) confidence limit of $N_{off}$, using Gaussian statistics. 
\begin{equation}
N_{on}^{LL\; 95}\approx N_{on}\times 
\left[1-{1\over 9N_{on}} - {1.645\over 3 \sqrt{N_{on}}}
+0.031N_{on}^{-2.50}\right]^3
\end{equation}
The above quantity is then the lower 95\% confidence limit for $N_{on}$ (eq. [3] of Mighell et 
al. 1998, and eq. [14] of Gehrels 1986). The field area was the same as the galaxy area in this case.
Then, in order to decide if any 
particular star is a cluster member, we generate a uniform random number, $0<p'<1$. If
$p>p'$, we accept the star as a member of the dSph. We ran the cleaning process twice, using the real off-galaxy 
field for the RGB stars, and then using the  TRILEGAL-generated objects for the whole sample. We
found that the results were consistent for the overalapping RGB stars. 

When the above process is performed for a globular cluster field with thousands (or tens of thousands) of stars, 
compared with a few hundred non-members, the statistical cleaning process rejects few real cluster
members. However, with the small numbers involved here, we are more careful to examine each object 
(which is also more feasible with 165 objects, rather than thousands). In Table~2, we have identified
classes of objects, A--F, which correspond with the likelihood of the object being a real dSph member.
Figure~5 is a finding chart for these objects based on the 300s R-image. Table~2 also lists the
J2000 equatorial coordinates of the stars, which we derived using the 2MASS stars in the image to 
construct the plate solution (using the IMCOORDS program suite in IRAF). 
Objects were ranked as class A if they passed the statistical cleaning process, were in the right color
area of Figure~4 (compared with the real off-galaxy field), and had photometry in all filters with
uncertainties less than 0.05.  A-class objects are shown as filled triangles in Figures 6 \& 7. 
B-class (open triangles) stars passed \it cleaning \rm process but had poorer quality
photometry (and are mostly at the MSTO). C-class objects (large open circles) failed the statistical cleaning process at the 
last stage when they were compared to the random probability, but had the right colors and good 
photometry. D-class stars (median-sized open circles) failed the statistical cleaning process, have the right colors but poor
photometry. E-class stars (small open circles) passed statistical cleaning but failed color-selection, and F-class stars (tiny open circles)
failed both statistical cleaning and color-selection.

We compared our RGB stars with the Martin et al. (2007) proper motion survey (see Table~2), and showed that we rejected the non-members in
the sample overlap. The only giant which fell outside the A-class was object \#8, classified C, which has the right colors, but was in a
region of the CMD with many foreground stars. We have shown that the statistical rejection method used alone is unlikely to select a
non-member, but might reject a few metal-poor RGB stars.

\section{Discussion \& Conclusions}

The $(T_1-T_2)$ vs.  $(C-T_1)$ color-color plot in Figure 6 of the A-class \boo\ stars (plus \#8) show that they
are of  low metallicity (mostly $[Fe/H]\leq -2.0$). The reddening here is very low, and does not affect the color-color
plot significantly. Distance modulus errors are more problematic for the CMDs, but the values used:
$(m-M)_V=19.11\pm 0.08\; mag$ and $E(B-V)=0.02$, give consistent fits to the isochrones in both
the CMDs and the color-color plots. The star with the very red $(T_1-T_2)$-color is \#29 in Table~2. This star sits at the base of the RGB in Figure~3,
and is likely to be a foreground dwarf that did not get rejected statistically.

The Geisler, Claria \& Minniti  (1991) lines of constant $[Fe/H]$ were derived from real RGB stars in globular and open clusters, where the metal-poor stars tend to be $\alpha$-enhanced. The models of Marigo et al. (2008)  calculate colors for MSTO and MS stars, but are solar-scaled. If we plotted the whole Z=0.0002 model in Figure~6, the MSTO would form a loop beneath the Geisler, Claria \& Minniti  (1991) grid, and the MS would cross the grid again, close to the locus of the RGB. We have MSTO stars in our data set, but the observations did not sample the MS stars. Any MS stars present on the color-color plots have to be metal-poor halo dwarfs. Since the error bars are relatively large compared to the spacing of the $[Fe/H]$-grid, we use the weighted mean of the [Fe/H]-values for the 7 brightest \boo\ members (filled squares),  $[Fe/H]= -2.1^{+0.3}_{-0.5}$. The range looks to be from $-3.5\leq  [Fe/H]\leq -1.8$, for these 7 RGB stars. The metal-rich \boo\ members appear to terminate at about -1.8, but the
the metal-poor stars may have lower values than [Fe/H]=-2.5. The Washington filter system  becomes less sensitive
to $[Fe/H]$ as the metals go to zero (see Figure 2b, where the fiducial RGBs get closer together in $(C-T_1)$ as $[Fe/H]$ reduces), so small errors in color can lead to large errors in $[Fe/H]$.

Our result is consistent with recently reported spectroscopy (Norris et al. 2008; Ivans et al. 2008). Norris et al. (2008) used the AAOmega multifibre facility on the Anglo-Australian Telescope, obtaining spectra over 3800--4600\,{\AA}.
They used the Ca II K line and the G-band to determine $[Fe/H] = -2.6\pm 0.5$ for 19 stars, which had S/N~$>$~14~per~0.34\,{\AA} pixel at 4100\,{\AA}. Ivans et al. (2008) used Magellan's multi-object echellette at Las Campanas Observatory to obtain higher resolution spectra of two \boo\ RGB stars, which were $[Fe/H]=-2.3$ (with $[\alpha /Fe]=+0.4$) and -1.9 ($[\alpha /Fe]=+0.2$). Considering the Salaris et al. (1993) formula for comparing 
solar-scaled to alpha-enhanced isochrones, $Z=Z_0(0.638f_\alpha + 0.362)$, where $Z_0$ is the non-enhanced metallicity and $f_\alpha$ is the
average enhancement factor, the RGB in Figure~7a would then appear to be $\geq 0.0001$ more metal-rich than is implied by Figure~6 (and the spectroscopic results).

We use the CMD and the color-color plot to estimate the metallicities of the stars, which both have their biasses.
 Figure~7a shows the CMD for the classes of objects from Table~2. We note that the
reddest RGB-tip star is also in the region with many foreground objects, whereas the MSTO stars
are relatively uncontaminated. The star (\#8) which is close to the horizontal branch area was not noted as
variable in Siegel (2006) and looks to be more metal rich that the rest of the class A stars, but is not inconsistent with the spectroscopic data, discussed
previously. The base
of the RGB is also an area which had many foreground contaminants. The relatively-clean MSTO
region shows a tight, single turn-off, which can be reproduced by the main \boo\ stars having
Z=0.0002 and an age of 14.1~Gyrs, with some stars being as metal-rich as Z=0.0003 at
12.2~Gyrs. We show that these isochrones (Marigo et al. 2008)\footnote{These new isochrones are solar-scaled, but there
has been considerable discussion in recent papers (see Bertelli et al. 2008; VandenBerg et al. 2007) on whether the new lower solar abundances
(Asplund et al. 2005; 2006) are inconsistent with helioseismology. Bertelli et al. use Z=0.017 for the sun, which they acknowledge is a 
compromise between the usual value of $Z\approx 0.02$ and the lower values derived by Asplund et al. (2005; 2006), but
matches the accepted  $R_\odot$, $L_\odot$ to within 0.2\% and
$R_c$ to about 1\%.}  overlap, giving a narrow MSTO-SGB
region, and are consistent with the color-color plot in Figure~6.

We note the presence of at least 4 blue straggler stars (BSS), which passed the cleaning process and are
well-separated from the other MSTO stars in $(T_1-T_2)$ and.  $(C-T_1)$. We show the isochrones
for the most metal-poor models (Z=0.0001) for 4.5 \& 0.004 Gyr. The bluest BSS seem to be on
the metal-poor main sequence, which appears to be more metal-poor than Z=0.0001. BSS are either
\it primordial \rm (having been formed with the rest of the stellar population in the system) or \it
collisional \rm (formed at different epochs due to collisions or close passes between stars). Momany
et al. (2007) discussed the BSS population of dSphs that had not undergone recent star formation, which included \boo .  They found that the BSS frequency for the lowest luminosity dwarf galaxies (which again includes \boo ) agreed 
with the frequency for the Milky Way halo and open clusters; and they derived a statistically significant 
$F^{BSS}_{HB}-M_V$ anti-correlation for these dSphs similar to that observed in globular clusters.
Figure~7 does show the suggestion of a \it blue plume, \rm which is an old BSS population (similar 
to that seen in open and globular clusters in our galaxy), but the photometric errors are larger, and
the color-separation is less clear.  In the A-class sample of Table~2, there is one
possible HB star and at least 4 BSS, making the Momany et al. (2007) statistic $F^{BSS}_{HB} = log(N_{BSS}/N_{HB})=0.6$, for this central region, whereas Momany et al. (2007) find it is $\sim 0.25$ for the
whole \boo\ field. The known RR Lyrae stars are outside our central field, but the central \boo\ population 
has its $F^{BSS}_{HB}$-value
closer to the mean BBS values for the Milky Way  halo (Preston \& Sneden 2000).  
Ferraro et al. (2006) analyzed the
BS population of $\omega$~Cen (a non-relaxed system), deciding that the stars had to be produced by non-collisional 
processes. Certainly, in this sparsely populated dSph, these BS also have to be non-collisional 
in origin. There is no evidence of recent star formation in \boo , with Bailin \& Ford (2007) finding
it devoid of HI gas.

 Unlike the previously-published photometry in $BVI$ 
(Siegel 2006; Dall'Ora et al. 2006) and $gi$ (B06), the MSTO region is clearly separated from the foreground stars in the Washington filters. Zucker et al. (2006b) saw clear evidence for a broadened MSTO and SGB in the UMa~II Dwarf, and while there is likely some spread in [Fe/H] at the MSTO, the  distribution of MSTO and SGB stars argues against a long period of continuous star formation (more than about 2 Gyr), and a $\geq 1.0$ dex spread in $[Fe/H]$.  Membership can be more properly determined by radial velocity studies, such as that by Mu\~{n}oz et al. (2006) and Martin et al. (2007),
but these can only be performed on large telescopes for giants. In the Mu\~{n}oz et al. (2006)
study (discussed in \S 1), 58 stars were observed, selected as having the correct colors (in the $gi$-filters) to be RGB or AGB objects belonging to \boo . Only 12 objects had radial velocities in the expected range ($95.6\pm 3.4\; km/s$), with only 7 being within the half-light radius , showing the need for having a better method of preselecting RGB candidates
before committing to spectroscopy. Our $CT_1T_2$  photometry  shows a narrow MSTO and SGB, 
with a  metallicity of $[Fe/H]=-2.1^{+0.3}_{-0.5}$ for the RGB and MSTO, with ages ranging from 12--14~Gyr, using the new Marigo et al. (2008) isochrones. In this study, \boo\ appears to be similar to, or slightly more metal poor than, M15, but does appear to have a small metallicity spread.

 \acknowledgements
The authors wish to thank  Ata Sarajedini and Doug Geisler for sharing their data, Ivan
King for advice. This paper used observations obtained with the Apache Point Observatory 3.5-meter telescope, which is owned and operated by the Astrophysical Research Consortium. 
Hughes is grateful for all the help provided by the APO telescope operators during remote observing.
We also acknowledge support from the Kennilworth Fund, of the New York Community Trust. Anne Bossi acknowledges support from the Washington NASA Space Grant Consortium and the Welch family.
We also thank Leo Girardi for providing the web interface for his isochrones and the TRILEGAL 1.1 code,
and an anonymous referee for improving the paper.

\clearpage
\centerline{\Large Figure Captions}

\bf Fig.1:- \rm Plot of the final uncertainties (calculated from the ALLSTAR uncertainty, aperture correction,  and the standard-star photometry, taken in quadrature) of the sources in Table 2. From the DAOPHOT-ALLSTAR output,
we selected objects with a CHI-value 
(the DAOPHOT goodness-of-fit statistic) between 0.5 
and 1.5. \bf (a) \rm The $\sigma_{T_1}$-value versus the $T_1$ magnitude.\\ \bf (b) \rm The uncertainty versus $T_1$ magnitude for $(C-T_1)$. \\ \bf (c) \rm The uncertainty versus $T_1$ magnitude for $(T_1-T_2).$

\bf Fig.2:- (a) \rm  $T_1$ vs.  $(C-T_1)$ CMD of the 165 stars detected in the \boo\ field, uncorrected for distance or extinction, shown as open squares with error bars.
 \\ \bf (b) \rm $M_{T_1}$ vs.  $(C-T_1)$ CMD of the stars in the \boo\  field. The small circles are M15 data from GS99.  The open squares with error bars are the 165 objects from Table 2, shown with the standard giant branches from GS99.  
We use the distance modulus of 19.11 and $E(B-V)=0.02$ for \boo ,
with $(m-M)_V=15.41$ and $E(B-V)=0.10$ for M15.

\bf Fig.3:- \rm $M_{T_1}$ vs.  $(C-T_1)$  CMD of the stars in the \boo\  field. The small circles are M15 data from GS99,  the open squares are the 165 objects from Table 2, the open stars are foreground stars from a nearby field, the crosses are a simulated stellar population using the TRILEGAL code (Girardi et al. 2005), using the same area and limiting magnitude. We plot the Z=0.0002, 14.1 Gyr, and Z=0.0006, 12.7 Gyr isochrones (Marigo et al. 2008) for comparison. For \boo , we use  
$(m-M)_V=19.11$ and $E(B-V)=0.02$.

\bf Fig.4:- \rm $(T_1 - T_2)$ vs.  $(C - T_1)$ color-color plot of the \boo\ stars with uncertainties better than 0.05, shown as filled squares. The 
constant-$[Fe/H]$ lines from Geisler, Claria \& Minniti  (1991) are marked.  The  Z=0.0006, 12.7 Gyr, isochrone of Marigo et al. (2008) is shown from the MSTO to the tip of the RGB.  Again, the open stars are foreground stars from a nearby field, the crosses are a simulated stellar population using the TRILEGAL code (Girardi et al. 2005), using the same area and limiting magnitude.

\bf Fig.5:- \rm The 300s R-band image of a field of the \boo\  dwarf galaxy, taken with the APO 3.5-m telescope and SPIcam (north is up, east is left. FOV$\sim 4.78^\prime \times 4.78^\prime$).  The numbers correspond to the sources in Table 2, with the plate solutions
from the 2MASS catalog to convert the xy-coordinates to right ascension and declination (J2000).

\bf Fig.6:- \rm  $(T_1-T_2)$ vs.  $(C-T_1)$ color-color plot of the \boo\ stars with uncertainties better than 0.05 in all filters and designated  class A (and
\#8, a C-class, radial velocity member) in Table 2, shown as filled triangles. Star \#19 is too red in $(T_1-T_2)$, and might be two unresolved stars. The constant-$[Fe/H]$ lines for RGB stars from Geisler, Claria \& Minniti  (1991) are marked. The  Z=0.0001, 0.0002, and 0.0003, 12.7 Gyr, isochrones of Marigo et al. (2008) are plotted from the MSTO to the tip of the RGB. We note that all the isochrones from 10-15 Gyr fall almost on top of each other on this color-color plot, since these colors are only sensitive to metallicity, not age. The 7 brightest cluster members are shown as filled squares.

\bf Fig.7:-  (a) \rm CMD for \boo\ stars. Filled triangles are class A, open triangles are class B, the classes C-F are decreasing sizes of open circles. For the filled triangles, the error bars are the same size as the points. We show various isochrones from Marigo et al. (2008), including those close to the possible blue stragglers. \\ \bf (b) \rm MSTO-SGB region of the CMD for \boo\ stars. Class A objects (filled triangles) have error bars, which are much larger on the other points and are not shown (class B, open triangles, are shown for their general trend). We show the  isochrones from Marigo et al. (2008).











\clearpage


\begin{deluxetable}{ccccc}
\tablenum{1}
\tablecolumns{5}
\tablewidth{0pc} 
\tabletypesize{\scriptsize}
\tablecaption{APO 3.5-m CCD Frames taken in Spring 2007}
\tablehead{
\colhead{Field} & \colhead{Filter}   &
 \colhead{Exposure(s)}  &    \colhead{Airmass$^1$}&   \colhead{FWHM(arcsec)$^2$}}
\startdata
\boo  $^3$& R& 1& 1.069& 0.9\\
\boo & R& 3& 1.067& 0.8\\
\boo & R& 10& 1.065& 0.8\\
\boo & R& 30& 1.064& 0.8\\
\boo & R& 90& 1.063& 0.8\\
\boo & R& 300& 1.060& 0.7\\
\boo & R& 1000& 1.058& 0.8\\
\boo & I& 1& 1.054& 0.6\\
\boo & I& 3& 1.053& 0.6\\
\boo & I& 10& 1.053& 0.6\\
\boo & I& 30& 1.053& 0.6\\
\boo & I& 90& 1.053& 0.7\\
\boo & I& 300& 1.053& 0.7\\
\boo & I& 1000& 1.054& 0.8\\
\boo & C& 1& 1.060& 0.7\\
\boo & C& 3& 1.061& 0.9\\
\boo & C& 10& 1.062& 0.8\\
\boo & C& 30& 1.063& 0.7\\
\boo & C& 90& 1.065& 0.8\\
\boo & C& 300& 1.068& 0.7\\
\boo & C& 1000& 1.072& 0.7\\
\enddata
\tablecomments{
(1) Effective airmass.
(2) Average seeing.
(3) 2007 March 19.}
\end{deluxetable}


\clearpage
 
 
\pagestyle{empty}
\begin{deluxetable}{lllllllllllll}
\rotate
\tablecolumns{13}
\tablewidth{0pc} 
\tabletypesize{\scriptsize}
\tablenum{2}
\tablecaption{Sample of Objects in Bootes I Central Field}
\tablehead{
\colhead{ID}&  \colhead{$X_C$} &  \colhead{$Y_C$}   &  \colhead{$T_1$} &  \colhead{$\sigma_{T_1}$} & \colhead{$C$} &  \colhead{$\sigma_{C}$} &  \colhead{$T_2$} &\colhead{$\sigma_{T_2}$} & \colhead{RA} & \colhead{Dec} & \colhead{Type}& \colhead{Note}}
\startdata
1& 465.650& 672.994& 11.364& 0.011& 12.868& 0.034& 10.900& 0.018& 14:00:07.26& 14:30:44.2& F& 2MASS 14000725+1430443\\
2& 404.362& 408.282& 14.334& 0.011& 15.575& 0.027& 13.957& 0.018& 14:00:08.43& 14:29:29.8& F& 2MASS 14000844+1429296\\
3& 960.096& 499.425& 16.011& 0.012& 19.482& 0.030& 14.356& 0.022& 13:59:57.69& 14:29:55.6& F& 2MASS 13595769+1429554, VR\\
4& 54.569& 53.073& 16.241& 0.011& 19.528& 0.033& 14.960& 0.026& 14:00:15.18& 14:27:49.8& F& 2MASS 14001516+1427499\\
5& 308.687& 801.328& 16.690& 0.011& 18.336& 0.027& 16.158& 0.018& 14:00:10.30& 14:31:20.2& F& 2MASS 14001029+1431205\\
6& 630.807& 818.253& 16.863& 0.011& 18.065& 0.026& 16.378& 0.018& 14:00:04.07& 14:31:25.1& E& 2MASS 14000406+1431251\\
7& 962.627& 499.406& 17.166& 0.014& 21.172& 0.061& 15.259& 0.023& 13:59:57.64& 14:29:55.6& E& \nodata \\
8& 298.684& 891.947& 17.224& 0.013& 19.050& 0.026& 16.612& 0.018& 14:00:10.49& 14:31:45.6& C& 2MASS 14001049+1431454, VA\\
9& 330.672& 171.268& 17.507& 0.011& 19.300& 0.028& 16.933& 0.017& 14:00:09.85& 14:28:23.1& A& 2MASS 14000985+1428228\\
10& 15.427& 605.009& 17.642& 0.011& 18.430& 0.030& 17.292& 0.017& 14:00:15.96& 14:30:24.9& F& \nodata \\
11& 891.801& 840.981& 17.712& 0.012& 20.972& 0.028& 16.608& 0.017& 13:59:59.02& 14:31:31.5& E& 2MASS 13595900+1431317\\
12& 480.032& 970.838& 17.771& 0.011& 19.350& 0.026& 17.305& 0.017& 14:00:06.99& 14:32:07.9& E& VR \\
13& 207.843& 915.000& 18.007& 0.011& 18.915& 0.029& 17.618& 0.017& 14:00:12.25& 14:31:52.1& F& 2MASS 14001226+1431518, VR\\
14& 540.963& 301.431& 18.626& 0.012& 19.588& 0.029& 18.230& 0.017& 14:00:05.78& 14:28:59.8& E& \nodata \\
15& 461.309& 912.008& 18.794& 0.011& 22.179& 0.033& 17.496& 0.017& 14:00:07.35& 14:31:51.3& E& 2MASS 14000735+1431511, VR\\
16& 287.259& 389.960& 18.919& 0.011& 20.289& 0.027& 18.401& 0.017& 14:00:10.69& 14:29:24.6& A& \nodata \\
17& 72.984& 337.386& 19.056& 0.012& 19.932& 0.029& 18.675& 0.018& 14:00:14.84& 14:29:09.7& F& \nodata \\
18& 171.206& 684.411& 19.228& 0.012& 21.339& 0.029& 18.639& 0.017& 14:00:12.95& 14:30:47.3& E& \nodata \\
19& 626.801& 6.284& 19.329& 0.017& 20.196& 0.030& 18.953& 0.018& 14:00:04.11& 14:27:36.9& A& Not RR Lyrae?\\
20& 329.809& 701.761& 19.370& 0.011& 22.101& 0.031& 18.513& 0.017& 14:00:09.88& 14:30:52.2& E& VR \\
21& 963.849& 2.411& 19.567& 0.013& 22.689& 0.057& 18.628& 0.019& 13:59:57.59& 14:27:35.9& E& \nodata \\
22& 950.939& 97.894& 19.776& 0.012& 20.970& 0.035& 19.302& 0.019& 13:59:57.85& 14:28:02.8& A& \nodata \\
23& 496.407& 265.649& 19.815& 0.012& 23.596& 0.054& 18.215& 0.017& 14:00:06.64& 14:28:49.7& E& \nodata \\
24& 667.937& 272.143& 20.127& 0.013& 21.189& 0.029& 19.626& 0.018& 14:00:03.33& 14:28:51.6& C& \nodata \\
25& 5.815& 897.782& 20.210& 0.012& 21.179& 0.034& 19.836& 0.020& 14:00:16.16& 14:31:47.1& F& \nodata \\
26& 589.614& 886.856& 20.301& 0.016& 23.082& 0.042& 19.417& 0.018& 14:00:04.87& 14:31:44.3& E& VR \\
27& 602.727& 828.935& 20.427& 0.013& 24.009& 0.056& 18.719& 0.018& 14:00:04.61& 14:31:28.1& E& VR \\
28& 564.802& 599.216& 20.472& 0.014& 21.576& 0.033& 19.968& 0.021& 14:00:05.34& 14:30:23.5& A& \nodata \\
29& 510.876& 714.457& 20.473& 0.015& 21.672& 0.030& 19.844& 0.022& 14:00:06.38& 14:30:55.8& A& \nodata \\
30& 672.201& 367.862& 20.570& 0.013& 24.163& 0.113& 19.064& 0.018& 14:00:03.25& 14:29:18.5& E& \nodata \\
31& 207.935& 381.417& 20.722& 0.015& 21.696& 0.030& 20.265& 0.021& 14:00:12.23& 14:29:22.1& C& \nodata \\
32& 692.683& 332.819& 20.775& 0.012& 21.539& 0.031& 20.414& 0.020& 14:00:02.85& 14:29:08.7& E& \nodata \\
33& 848.414& 14.943& 20.845& 0.014& 21.808& 0.038& 20.481& 0.025& 13:59:59.83& 14:27:39.4& E& Close to V8\\
34& 681.498& 600.224& 20.870& 0.014& 21.992& 0.038& 20.407& 0.020& 14:00:03.08& 14:30:23.8& A& VA \\
35& 716.246& 951.889& 20.894& 0.021& 23.482& 0.054& 20.188& 0.021& 14:00:02.42& 14:32:02.6& E& \nodata \\
36& 653.274& 903.421& 20.954& 0.016& 22.070& 0.041& 20.546& 0.020& 14:00:03.64& 14:31:49.0& E& \nodata \\
37& 925.807& 867.628& 21.055& 0.014& 22.859& 0.044& 20.465& 0.020& 13:59:58.36& 14:31:39.0& E& \nodata \\
38& 711.904& 758.333& 21.081& 0.061& 23.262& 0.084& 20.340& 0.050& 14:00:02.50& 14:31:08.3& F& \nodata \\
39& 664.767& 875.856& 21.264& 0.016& 23.105& 0.044& 20.655& 0.024& 14:00:03.41& 14:31:41.3& F& \nodata \\
40& 789.328& 824.543& 21.285& 0.017& 22.411& 0.035& 20.780& 0.025& 14:00:01.00& 14:31:26.9& A& \nodata \\
41& 641.611& 905.460& 21.377& 0.055& 23.642& 0.085& 20.705& 0.053& 14:00:03.86& 14:31:49.6& E& \nodata \\
42& 917.939& 724.559& 21.486& 0.014& 22.618& 0.035& 21.083& 0.038& 13:59:58.51& 14:30:58.8& C& \nodata \\
43& 6.632& 97.070& 21.507& 0.016& 22.604& 0.052& 21.216& 0.034& 14:00:16.11& 14:28:02.1& E& \nodata \\
44& 945.510& 434.384& 21.521& 0.019& 22.518& 0.035& 21.128& 0.024& 13:59:57.97& 14:29:37.3& E& \nodata \\
45& 545.050& 475.098& 21.539& 0.015& 21.745& 0.031& 20.970& 0.031& 14:00:05.71& 14:29:48.6& A& BSS\\
46& 452.631& 123.614& 21.728& 0.021& 23.565& 0.060& 21.142& 0.023& 14:00:07.48& 14:28:09.8& E& \nodata \\
47& 673.778& 935.680& 21.751& 0.023& 22.669& 0.037& 21.282& 0.032& 14:00:03.24& 14:31:58.1& A& \nodata \\
48& 845.796& 760.066& 21.817& 0.019& 21.733& 0.033& 21.650& 0.042& 13:59:59.91& 14:31:08.8& A& BSS\\
49& 361.348& 389.829& 21.832& 0.016& 24.245& 0.087& 20.416& 0.019& 14:00:09.26& 14:29:24.6& E& \nodata \\
50& 87.948& 538.212& 21.915& 0.019& 22.771& 0.041& 21.513& 0.028& 14:00:14.56& 14:30:06.1& A& \nodata \\
51& 661.843& 801.672& 21.921& 0.023& 22.252& 0.035& 21.640& 0.043& 14:00:03.47& 14:31:20.4& A& BSS\\
52& 137.089& 649.530& 21.940& 0.023& 22.829& 0.040& 21.519& 0.040& 14:00:13.61& 14:30:37.4& A& \nodata \\
53& 977.369& 783.827& 21.946& 0.022& 22.597& 0.042& 21.529& 0.057& 13:59:57.36& 14:31:15.5& B& \nodata \\
54& 148.239& 789.076& 21.971& 0.022& 22.848& 0.042& 21.504& 0.048& 14:00:13.40& 14:31:16.7& A& \nodata \\
55& 218.112& 626.465& 21.980& 0.084& 25.480& 0.295& 20.289& 0.019& 14:00:12.04& 14:30:31.0& E& \nodata \\
56& 957.905& 376.235& 21.982& 0.025& 22.820& 0.041& 21.660& 0.055& 13:59:57.72& 14:29:21.0& B& \nodata \\
57& 138.957& 212.847& 22.045& 0.031& 22.464& 0.053& 21.844& 0.044& 14:00:13.55& 14:28:34.7& B& BSS\\
58& 908.028& 210.687& 22.047& 0.075& 24.180& 0.342& 21.442& 0.040& 13:59:58.68& 14:28:34.4& F& \nodata \\
59& 189.245& 603.444& 22.064& 0.022& 22.705& 0.059& 21.781& 0.052& 14:00:12.60& 14:30:24.5& E& \nodata \\
60& 676.467& 342.658& 22.066& 0.018& 22.843& 0.042& 21.635& 0.058& 14:00:03.16& 14:29:11.4& B& \nodata \\
61& 957.068& 756.913& 22.067& 0.023& 22.912& 0.041& 21.595& 0.039& 13:59:57.75& 14:31:07.9& A& \nodata \\
62& 979.011& 569.678& 22.096& 0.016& 22.912& 0.041& 21.832& 0.055& 13:59:57.32& 14:30:15.3& E& \nodata \\
63& 619.808& 1012.61& 22.144& 0.017& 22.848& 0.044& 21.821& 0.035& 14:00:04.29& 14:32:19.7& A& \nodata \\
64& 536.738& 928.120& 22.146& 0.028& 22.898& 0.039& 21.678& 0.048& 14:00:05.89& 14:31:55.9& A& \nodata \\
65& 974.000& 568.127& 22.195& 0.026& 22.913& 0.049& 21.987& 0.042& 13:59:57.42& 14:30:14.9& F& \nodata \\
66& 33.712& 77.835& 22.197& 0.042& 22.988& 0.050& 21.614& 0.038& 14:00:15.58& 14:27:56.8& B& \nodata \\
67& 341.736& 365.060& 22.201& 0.025& 22.283& 0.035& 22.227& 0.048& 14:00:09.64& 14:29:17.6& A& \nodata \\
68& 952.338& 152.183& 22.226& 0.024& 22.700& 0.044& 22.098& 0.080& 13:59:57.82& 14:28:18.0& B& \nodata \\
69& 956.856& 124.497& 22.278& 0.029& 23.263& 0.048& 21.875& 0.065& 13:59:57.73& 14:28:10.2& E& \nodata \\
70& 426.090& 967.923& 22.334& 0.039& 23.146& 0.047& 21.927& 0.033& 14:00:08.03& 14:32:07.0& A& \nodata \\
71& 810.443& 896.079& 22.334& 0.028& 23.113& 0.043& 21.722& 0.061& 14:00:00.60& 14:31:47.0& B& \nodata \\
72& 882.343& 409.180& 22.342& 0.019& 23.079& 0.050& 21.953& 0.037& 13:59:59.19& 14:29:30.2& B& \nodata \\
73& 148.738& 278.495& 22.381& 0.026& 23.923& 0.109& 21.713& 0.038& 14:00:13.37& 14:28:53.2& E& \nodata \\
74& 253.582& 998.234& 22.395& 0.021& 23.613& 0.058& 21.939& 0.042& 14:00:11.37& 14:32:15.5& E& \nodata \\
75& 958.295& 953.537& 22.407& 0.025& 23.169& 0.056& 22.136& 0.079& 13:59:57.74& 14:32:03.2& B& \nodata \\
76& 77.620& 796.977& 22.409& 0.019& 23.252& 0.050& 22.353& 0.048& 14:00:14.77& 14:31:18.9& E& \nodata \\
77& 820.070& 745.891& 22.426& 0.052& 22.975& 0.044& 22.008& 0.053& 14:00:00.40& 14:31:04.8& B& \nodata \\
78& 150.535& 411.135& 22.449& 0.026& 23.113& 0.047& 22.252& 0.075& 14:00:13.34& 14:29:30.5& E& \nodata \\
79& 57.414& 71.384& 22.477& 0.020& 23.350& 0.071& 21.924& 0.051& 14:00:15.13& 14:27:55.0& E& \nodata \\
80& 587.436& 178.073& 22.481& 0.035& 23.009& 0.047& 22.044& 0.064& 14:00:04.88& 14:28:25.2& B& \nodata \\
81& 372.655& 921.741& 22.496& 0.019& 23.161& 0.051& 22.293& 0.108& 14:00:09.07& 14:31:54.0& B& \nodata \\
82& 313.526& 867.753& 22.515& 0.019& 23.258& 0.054& 22.298& 0.090& 14:00:10.21& 14:31:38.8& B& \nodata \\
83& 243.621& 710.867& 22.519& 0.032& 23.243& 0.056& 22.403& 0.066& 14:00:11.55& 14:30:54.7& E& \nodata \\
84& 334.968& 513.434& 22.523& 0.028& 23.079& 0.054& 22.062& 0.053& 14:00:09.78& 14:29:59.3& B& \nodata \\
85& 226.373& 789.191& 22.553& 0.035& 23.252& 0.049& 22.282& 0.048& 14:00:11.89& 14:31:16.7& A& \nodata \\
86& 77.601& 787.030& 22.566& 0.022& 23.308& 0.052& 22.236& 0.042& 14:00:14.77& 14:31:16.1& B& \nodata \\
87& 406.006& 57.568& 22.588& 0.029& 23.212& 0.054& 22.363& 0.050& 14:00:08.38& 14:27:51.2& E& \nodata \\
88& 858.335& 565.959& 22.592& 0.031& 23.185& 0.055& 22.247& 0.101& 13:59:59.66& 14:30:14.3& B& \nodata \\
89& 613.419& 448.395& 22.601& 0.030& 23.063& 0.064& 22.277& 0.075& 14:00:04.39& 14:29:41.1& D& \nodata \\
90& 201.889& 449.136& 22.605& 0.038& 22.984& 0.047& 22.599& 0.054& 14:00:12.35& 14:29:41.2& B& \nodata \\
91& 385.872& 256.560& 22.638& 0.039& 23.085& 0.047& 22.202& 0.060& 14:00:08.78& 14:28:47.1& B& \nodata \\
92& 324.001& 760.919& 22.650& 0.023& 23.456& 0.067& 22.245& 0.075& 14:00:10.00& 14:31:08.8& B& \nodata \\
93& 753.361& 54.971& 22.656& 0.037& 23.434& 0.062& 22.282& 0.101& 14:00:01.67& 14:27:50.6& B& \nodata \\
94& 610.504& 506.923& 22.670& 0.021& 23.225& 0.054& 22.466& 0.091& 14:00:04.45& 14:29:57.6& E& \nodata \\
95& 280.485& 123.461& 22.675& 0.025& 23.360& 0.050& 22.236& 0.047& 14:00:10.81& 14:28:09.7& B& \nodata \\
96& 483.257& 776.163& 22.675& 0.029& 23.969& 0.092& 21.783& 0.036& 14:00:06.92& 14:31:13.2& E& \nodata \\
97& 760.545& 325.963& 22.699& 0.026& 23.756& 0.084& 21.869& 0.035& 14:00:01.54& 14:29:06.8& E& \nodata \\
98& 311.522& 242.684& 22.699& 0.024& 23.178& 0.050& 22.416& 0.065& 14:00:10.22& 14:28:43.2& B& \nodata \\
99& 492.600& 869.320& 22.721& 0.022& 23.405& 0.092& 22.468& 0.075& 14:00:06.74& 14:31:39.4& B& \nodata \\
100& 182.199& 285.789& 22.734& 0.033& 23.554& 0.058& 22.371& 0.057& 14:00:12.72& 14:28:55.3& E& \nodata \\
101& 239.042& 3.639& 22.743& 0.045& 23.467& 0.058& 22.882& 0.142& 14:00:11.61& 14:27:36.0& E& \nodata \\
102& 952.695& 130.212& 22.757& 0.038& 23.439& 0.052& 22.287& 0.048& 13:59:57.81& 14:28:11.8& B& \nodata \\
103& 943.911& 485.903& 22.773& 0.035& 23.733& 0.079& 21.413& 0.103& 13:59:58.00& 14:29:51.8& E& \nodata \\
104& 878.449& 266.052& 22.785& 0.031& 23.670& 0.154& 22.100& 0.059& 13:59:59.26& 14:28:50.0& B& \nodata \\
105& 480.903& 322.344& 22.792& 0.024& 23.464& 0.061& 22.571& 0.091& 14:00:06.95& 14:29:05.6& B& \nodata \\
106& 420.243& 500.004& 22.797& 0.047& 23.343& 0.076& 22.638& 0.135& 14:00:08.13& 14:29:55.5& B& \nodata \\
107& 694.793& 586.227& 22.805& 0.042& 23.360& 0.067& 22.288& 0.052& 14:00:02.82& 14:30:19.9& B& \nodata \\
108& 40.007& 597.012& 22.810& 0.041& 23.408& 0.100& 22.413& 0.109& 14:00:15.49& 14:30:22.6& B& \nodata \\
109& 343.575& 849.542& 22.811& 0.028& 23.425& 0.058& 22.395& 0.054& 14:00:09.62& 14:31:33.7& B& \nodata \\
110& 641.122& 456.282& 22.813& 0.040& 23.219& 0.064& 22.413& 0.045& 14:00:03.85& 14:29:43.3& B& \nodata \\
111& 279.935& 508.314& 22.814& 0.031& 23.953& 0.104& 21.846& 0.034& 14:00:10.84& 14:29:57.8& E& \nodata \\
112& 908.838& 495.568& 22.816& 0.048& 23.313& 0.051& 22.252& 0.073& 13:59:58.68& 14:29:54.5& B& \nodata \\
113& 402.070& 893.008& 22.828& 0.042& 23.504& 0.062& 22.386& 0.078& 14:00:08.50& 14:31:46.0& B& \nodata \\
114& 716.162& 994.159& 22.846& 0.040& 23.794& 0.076& 22.341& 0.089& 14:00:02.42& 14:32:14.5& E& \nodata \\
115& 121.874& 698.918& 22.850& 0.043& 23.437& 0.062& 22.312& 0.084& 14:00:13.91& 14:30:51.3& B& \nodata \\
116& 836.493& 275.678& 22.853& 0.039& 23.410& 0.055& 22.784& 0.126& 14:00:00.07& 14:28:52.7& E& \nodata \\
117& 599.412& 853.925& 22.866& 0.025& 23.445& 0.058& 22.944& 0.135& 14:00:04.68& 14:31:35.1& E& \nodata \\
118& 738.943& 659.646& 22.873& 0.053& 23.457& 0.080& 22.592& 0.091& 14:00:01.97& 14:30:40.5& B& \nodata \\
119& 558.704& 834.797& 22.875& 0.025& 23.414& 0.068& 22.514& 0.056& 14:00:05.46& 14:31:29.7& B& \nodata \\
120& 411.578& 218.014& 22.879& 0.040& 23.488& 0.080& 22.579& 0.116& 14:00:08.28& 14:28:36.3& B& \nodata \\
121& 505.004& 301.975& 22.912& 0.042& 23.649& 0.072& 22.387& 0.099& 14:00:06.48& 14:28:59.9& B& \nodata \\
122& 179.055& 455.900& 22.926& 0.026& 23.641& 0.097& 22.580& 0.090& 14:00:12.79& 14:29:43.1& B& \nodata \\
123& 730.465& 611.611& 22.927& 0.027& 23.572& 0.077& 22.460& 0.100& 14:00:02.13& 14:30:27.0& B& \nodata \\
124& 862.889& 929.290& 22.928& 0.045& 23.558& 0.053& 22.729& 0.069& 13:59:59.58& 14:31:56.3& E& \nodata \\
125& 608.254& 752.400& 22.957& 0.053& 23.729& 0.083& 22.343& 0.083& 14:00:04.50& 14:31:06.5& B& \nodata \\
126& 786.331& 979.146& 22.961& 0.047& 23.785& 0.071& 22.465& 0.094& 14:00:01.07& 14:32:10.3& B& \nodata \\
127& 953.222& 991.316& 22.976& 0.042& 23.819& 0.073& 22.445& 0.064& 13:59:57.84& 14:32:13.8& B& \nodata \\
128& 665.141& 568.773& 22.985& 0.049& 23.724& 0.074& 22.997& 0.095& 14:00:03.39& 14:30:15.0& E& \nodata \\
129& 963.374& 829.092& 23.017& 0.062& 23.879& 0.072& 23.113& 0.107& 13:59:57.63& 14:31:28.2& E& \nodata \\
130& 807.932& 279.447& 23.025& 0.060& 23.526& 0.061& 22.511& 0.082& 14:00:00.62& 14:28:53.7& B& \nodata \\
131& 399.059& 356.671& 23.038& 0.028& 23.594& 0.072& 22.509& 0.122& 14:00:08.53& 14:29:15.3& B& \nodata \\
132& 898.085& 163.200& 23.051& 0.067& 23.651& 0.074& 22.476& 0.054& 13:59:58.87& 14:28:21.1& B& \nodata \\
133& 967.446& 717.057& 23.074& 0.038& 23.484& 0.077& 22.334& 0.059& 13:59:57.55& 14:30:56.7& B& \nodata \\
134& 229.554& 516.635& 23.079& 0.027& 23.796& 0.093& 22.512& 0.094& 14:00:11.82& 14:30:00.1& B& \nodata \\
135& 728.910& 249.357& 23.081& 0.085& 23.810& 0.103& 22.449& 0.112& 14:00:02.15& 14:28:45.2& B& \nodata \\
136& 532.350& 73.423& 23.095& 0.052& 23.696& 0.075& 22.636& 0.092& 14:00:05.94& 14:27:55.7& B& \nodata \\
137& 777.719& 982.319& 23.121& 0.033& 23.878& 0.097& 22.588& 0.067& 14:00:01.23& 14:32:11.2& B& \nodata \\
138& 577.228& 1009.81& 23.132& 0.050& 23.801& 0.081& 22.728& 0.074& 14:00:05.11& 14:32:18.9& B& \nodata \\
139& 505.335& 239.159& 23.190& 0.057& 23.787& 0.086& 22.752& 0.080& 14:00:06.47& 14:28:42.3& B& \nodata \\
140& 895.176& 649.082& 23.198& 0.069& 23.667& 0.070& 23.163& 0.116& 13:59:58.95& 14:30:37.6& B& \nodata \\
141& 498.063& 375.604& 23.235& 0.080& 23.812& 0.092& 22.920& 0.142& 14:00:06.62& 14:29:20.6& B& \nodata \\
142& 10.353& 402.258& 23.242& 0.060& 23.957& 0.077& 22.582& 0.134& 14:00:16.05& 14:29:27.9& B& \nodata \\
143& 201.894& 479.908& 23.266& 0.067& 23.983& 0.107& 22.234& 0.065& 14:00:12.35& 14:29:49.8& B& \nodata \\
144& 342.190& 368.437& 23.277& 0.069& 23.893& 0.111& 23.459& 0.174& 14:00:09.63& 14:29:18.5& E& \nodata \\
145& 252.969& 910.391& 23.281& 0.066& 23.769& 0.073& 23.341& 0.135& 14:00:11.38& 14:31:50.8& D& \nodata \\
146& 451.268& 822.990& 23.296& 0.039& 23.468& 0.070& 22.979& 0.092& 14:00:07.54& 14:31:26.3& B& \nodata \\
147& 7.398& 1008.48& 23.311& 0.060& 24.098& 0.094& 23.808& 0.247& 14:00:16.13& 14:32:18.3& E& \nodata \\
148& 905.426& 391.313& 23.312& 0.064& 24.045& 0.117& 22.733& 0.136& 13:59:58.74& 14:29:25.2& B& \nodata \\
149& 678.311& 200.695& 23.356& 0.067& 23.672& 0.077& 23.346& 0.133& 14:00:03.12& 14:28:31.5& B& \nodata \\
150& 718.523& 463.654& 23.362& 0.039& 24.077& 0.106& 22.908& 0.129& 14:00:02.36& 14:29:45.5& B& \nodata \\
151& 828.898& 602.344& 23.367& 0.037& 24.145& 0.109& 22.580& 0.059& 14:00:00.23& 14:30:24.5& B& \nodata \\
152& 199.140& 888.963& 23.384& 0.049& 23.989& 0.091& 23.028& 0.100& 14:00:12.42& 14:31:44.8& D& \nodata \\
153& 26.274& 659.335& 23.394& 0.067& 24.001& 0.105& 22.582& 0.066& 14:00:15.75& 14:30:40.2& B& \nodata \\
154& 404.574& 911.300& 23.417& 0.078& 23.990& 0.103& 23.547& 0.161& 14:00:08.45& 14:31:51.1& E& \nodata \\
155& 697.135& 261.617& 23.471& 0.069& 23.926& 0.100& 22.560& 0.065& 14:00:02.76& 14:28:48.7& B& \nodata \\
156& 582.196& 364.987& 23.475& 0.072& 24.125& 0.126& 23.432& 0.141& 14:00:04.99& 14:29:17.7& E& \nodata \\
157& 377.303& 345.663& 23.489& 0.067& 24.127& 0.100& 23.273& 0.121& 14:00:08.95& 14:29:12.2& B& \nodata \\
158& 311.717& 690.271& 23.506& 0.050& 24.219& 0.122& 23.946& 0.296& 14:00:10.23& 14:30:49.0& E& \nodata \\
159& 657.320& 652.457& 23.517& 0.043& 24.127& 0.107& 22.865& 0.089& 14:00:03.55& 14:30:38.5& B& \nodata \\
160& 237.679& 977.635& 23.536& 0.078& 24.366& 0.121& 22.767& 0.082& 14:00:11.68& 14:32:09.7& B& \nodata \\
161& 688.043& 796.889& 23.547& 0.045& 24.144& 0.121& 22.515& 0.134& 14:00:02.96& 14:31:19.1& B& \nodata \\
162& 522.233& 950.515& 23.559& 0.073& 24.287& 0.107& 22.885& 0.109& 14:00:06.17& 14:32:02.2& B& \nodata \\
163& 270.179& 463.211& 23.579& 0.049& 24.265& 0.150& 22.477& 0.056& 14:00:11.03& 14:29:45.1& B& \nodata \\
164& 471.906& 883.595& 23.605& 0.047& 24.321& 0.126& 23.996& 0.240& 14:00:07.14& 14:31:43.4& E& \nodata \\
165& 576.388& 537.281& 23.636& 0.044& 24.085& 0.104& 23.354& 0.134& 14:00:05.11& 14:30:06.1& B& \nodata \\
\enddata
\tablecomments{\\
A - Sources which have passed statistical cleaning and color-selection, and which have uncertainties better than 0.05 in all filters.\\
B - Sources which have passed statistical cleaning and color-selection, which do not have uncertainties better than 0.05 in all filters.\\
C - Sources which passed color-selection  failed statistical cleaning, and which have uncertainties better than 0.05 in all filters.\\
D - Sources which passed  color-selection, failed statistical cleaning, and do not have uncertainties better than 0.05 in all filters.\\
E - Sources which passed statistical cleaning but failed color selection. \\
F - Sources which failed statistical cleaning and color selection.\\
VA - Source accepted by Martin et al.'s (2006) radial velocity study.\\
VR - Source rejected by Martin et al.'s (2006) radial velocity study. \\
BSS - Possible blue straggler star.}
\end{deluxetable}


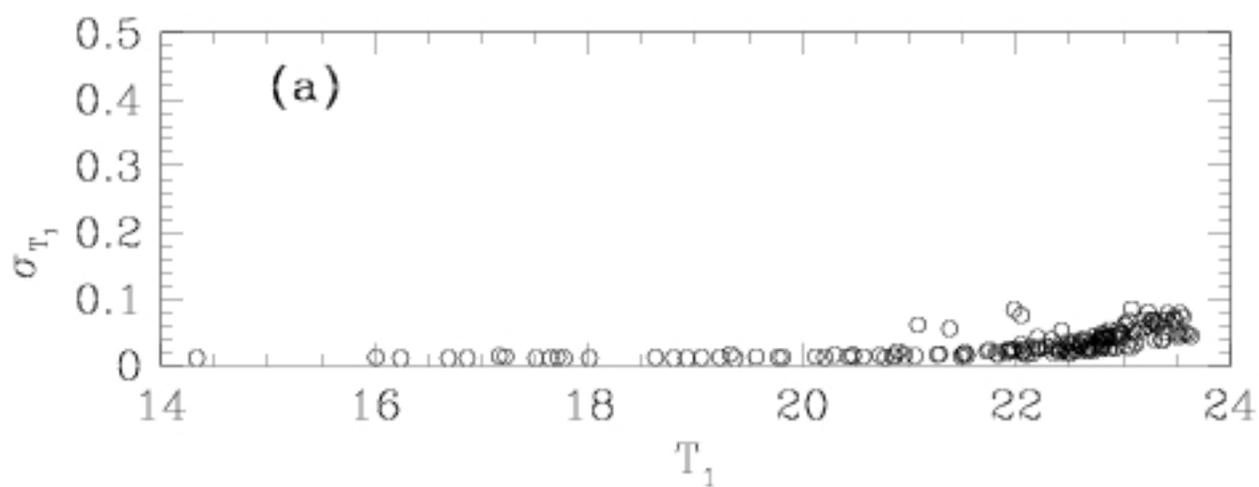
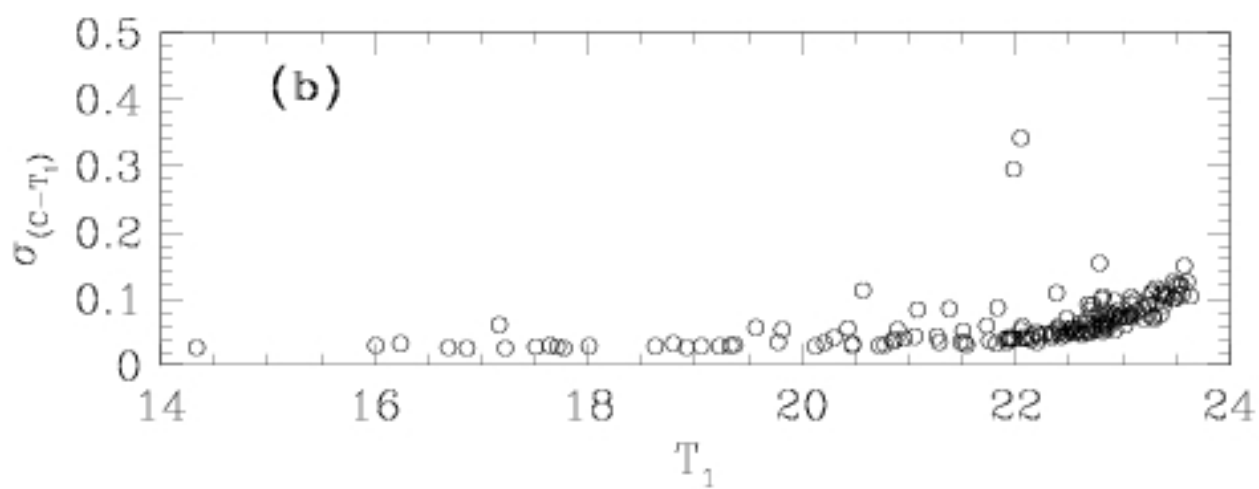
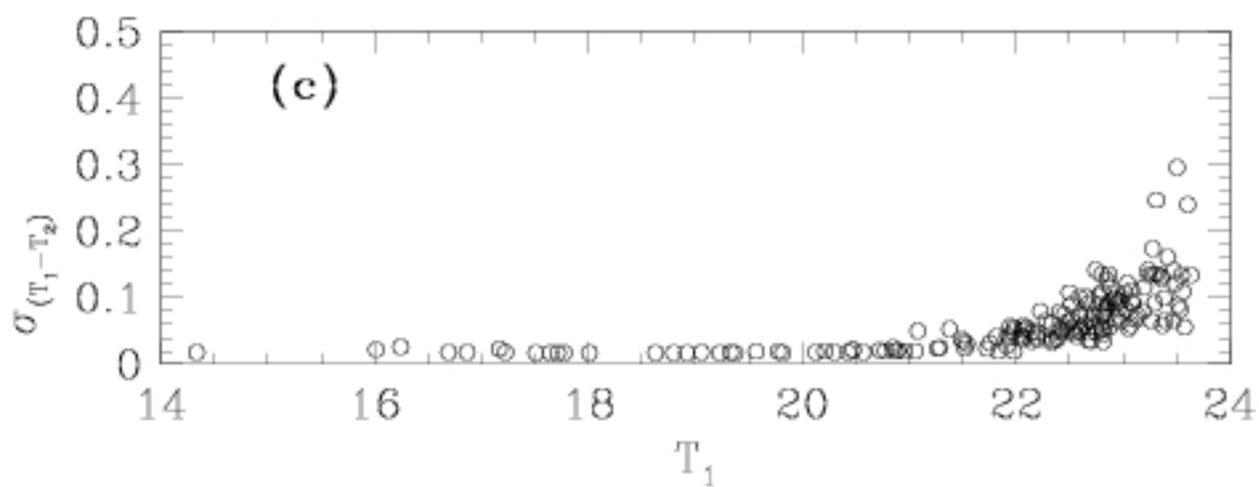
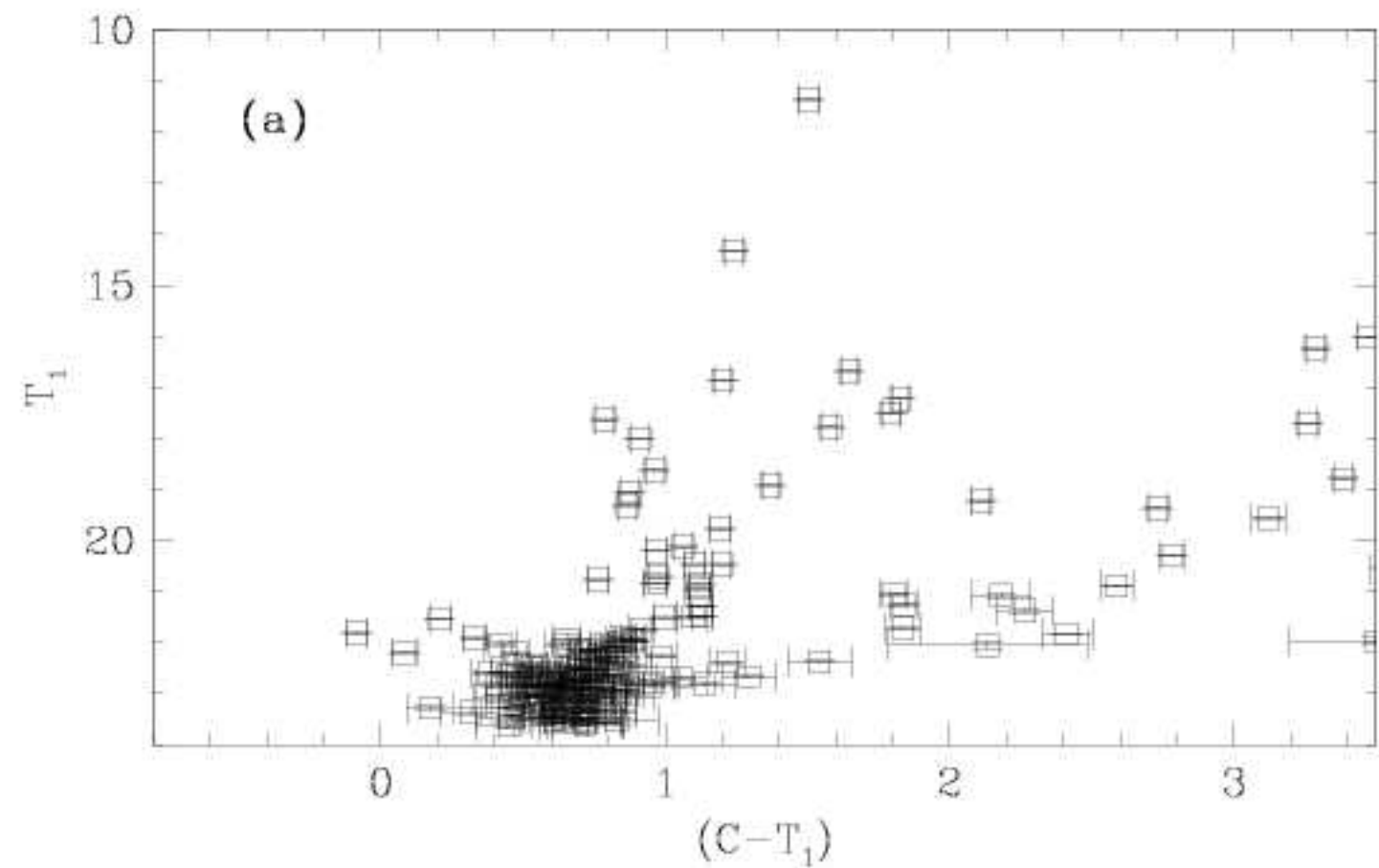
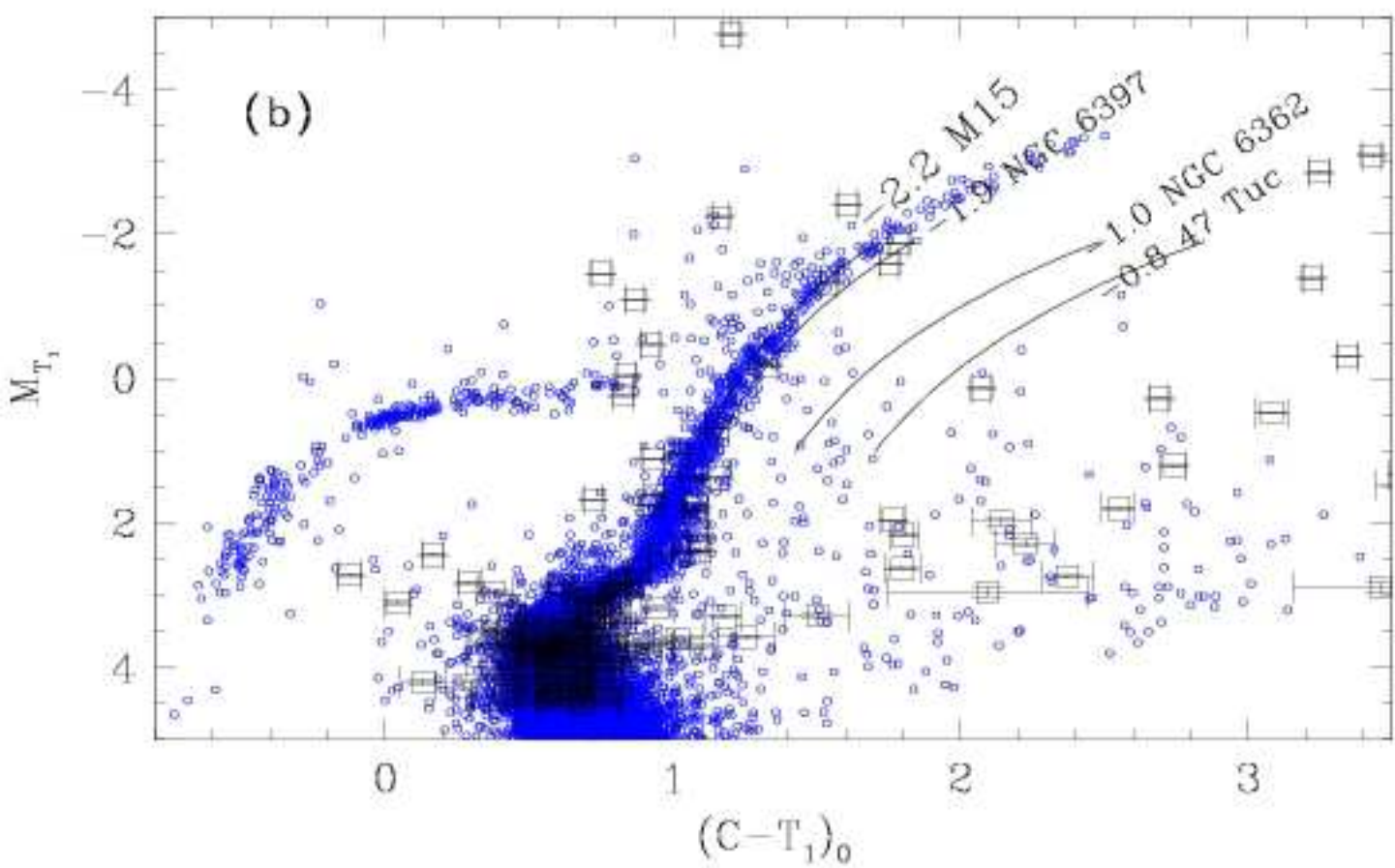
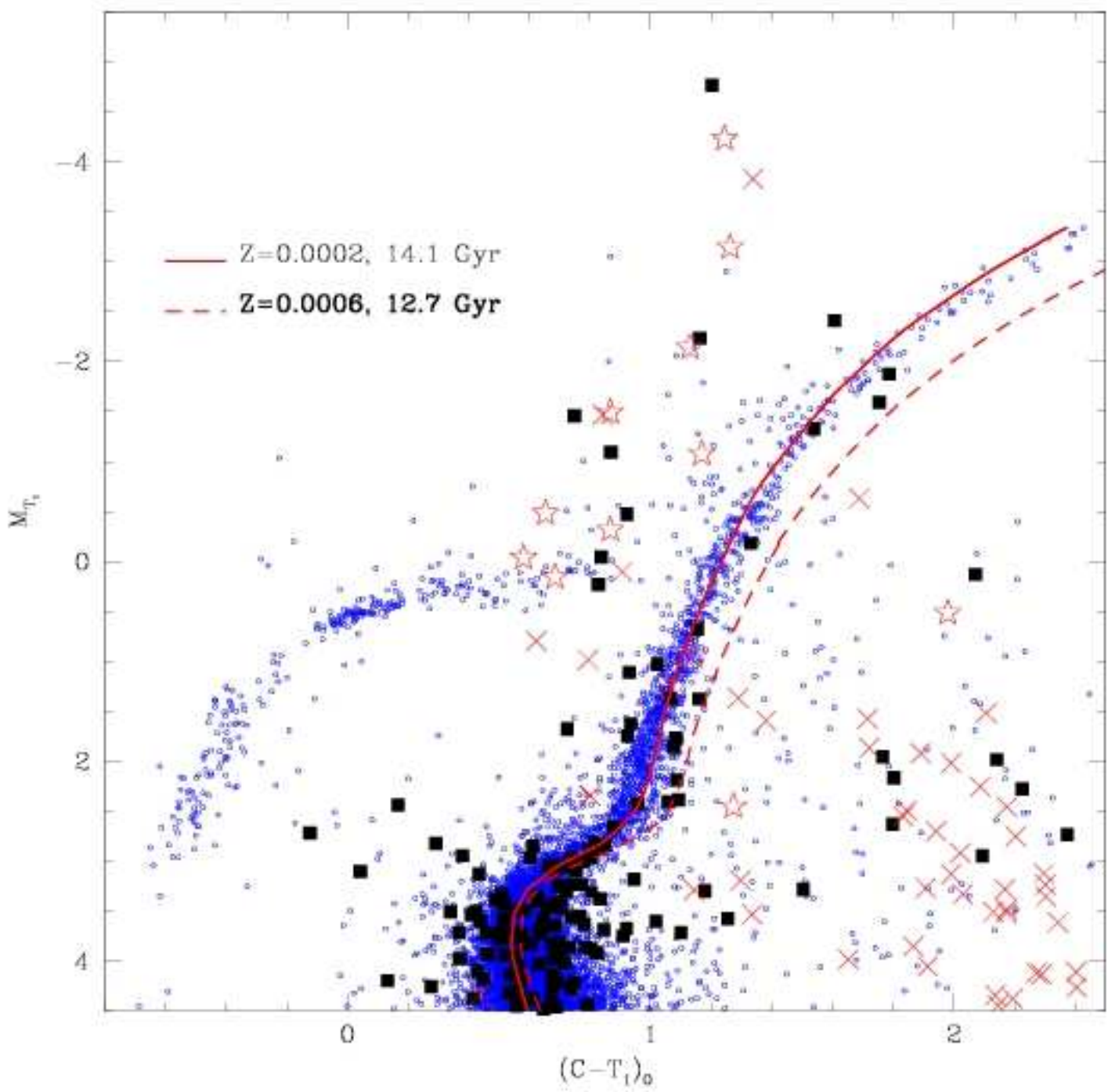
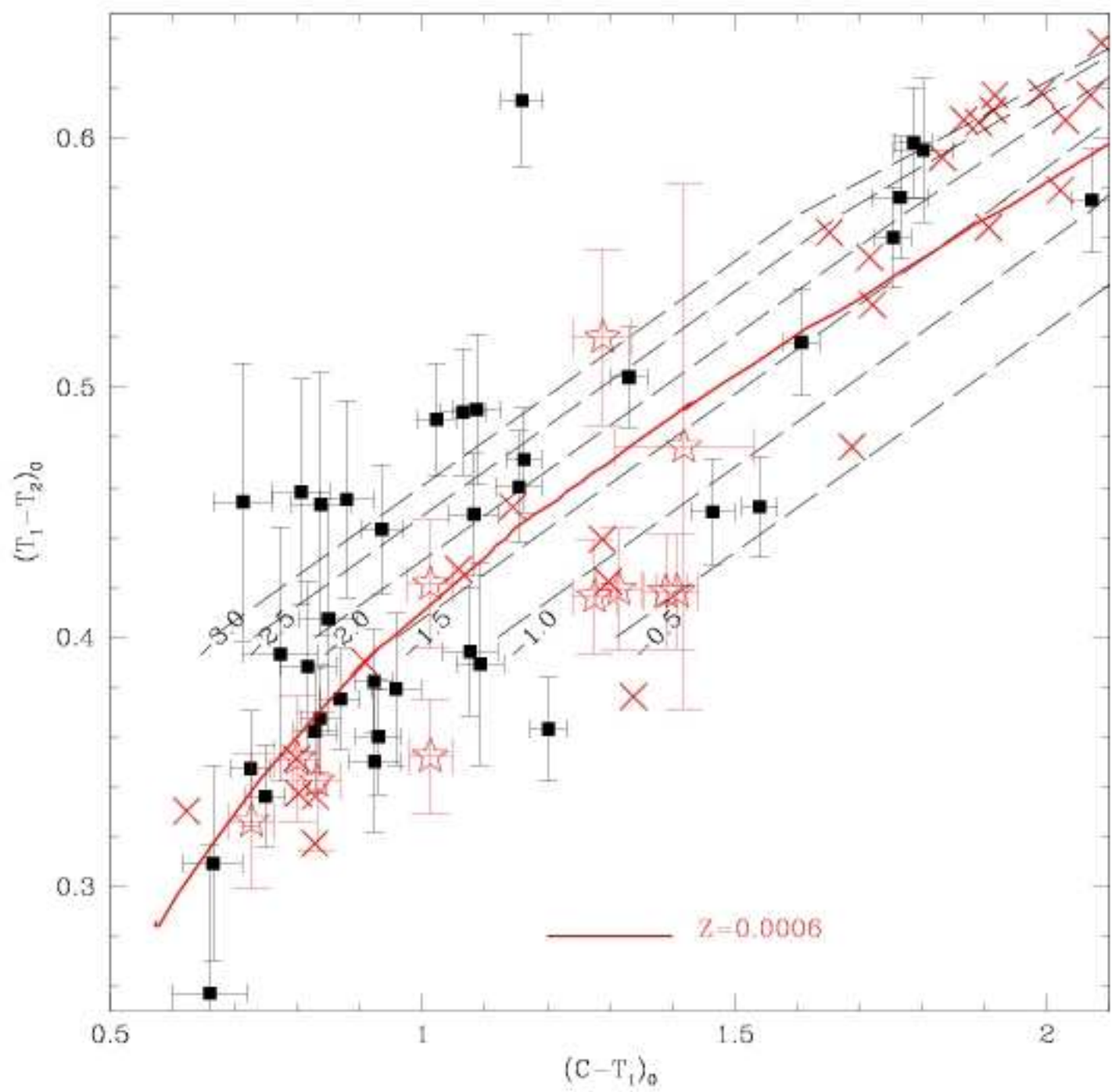
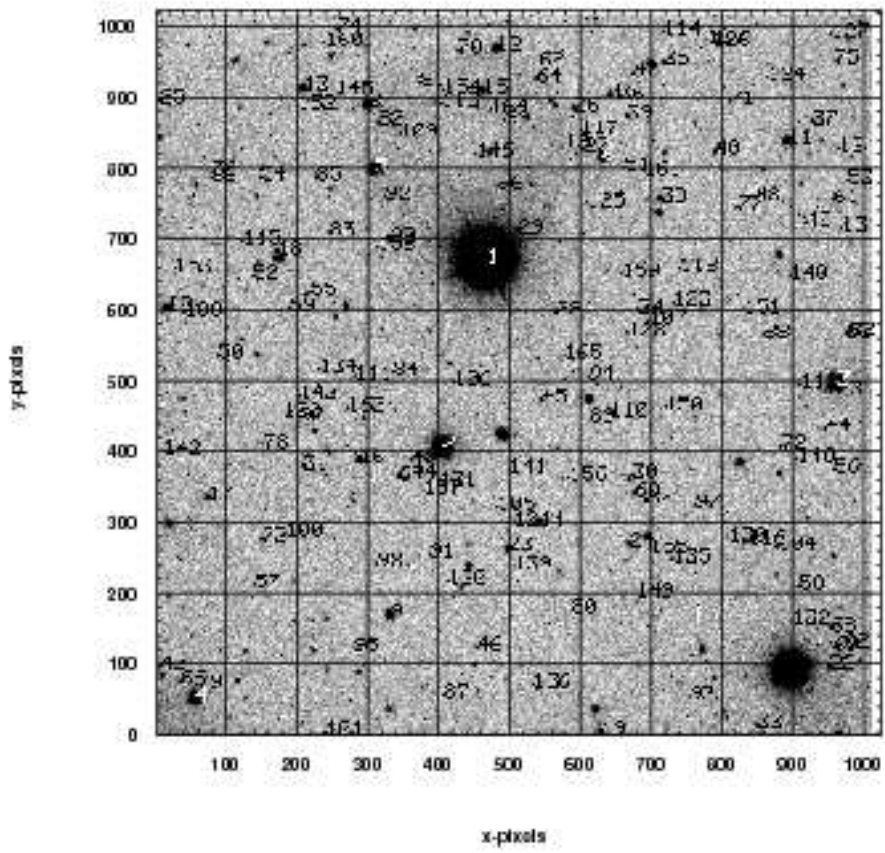
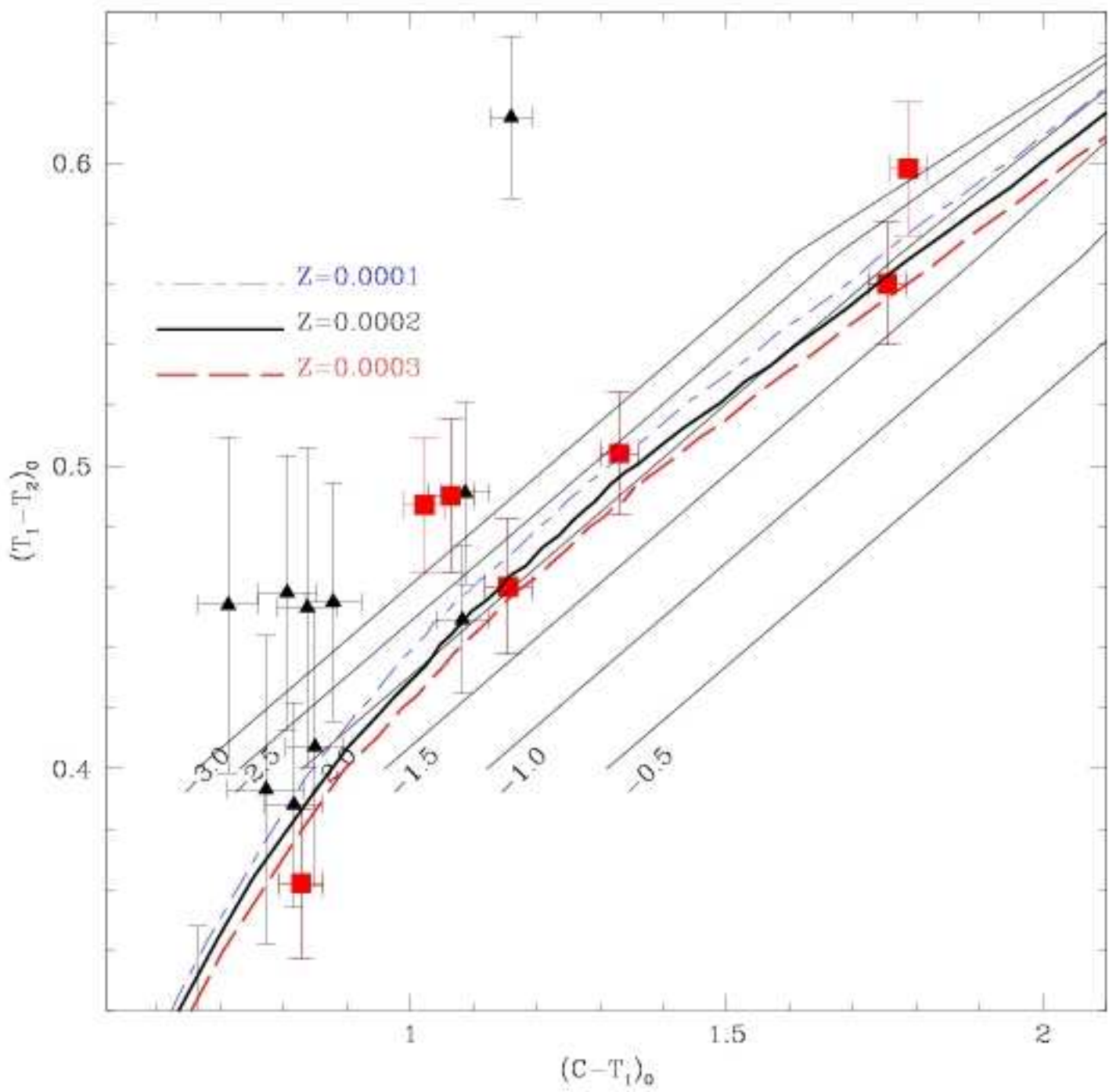
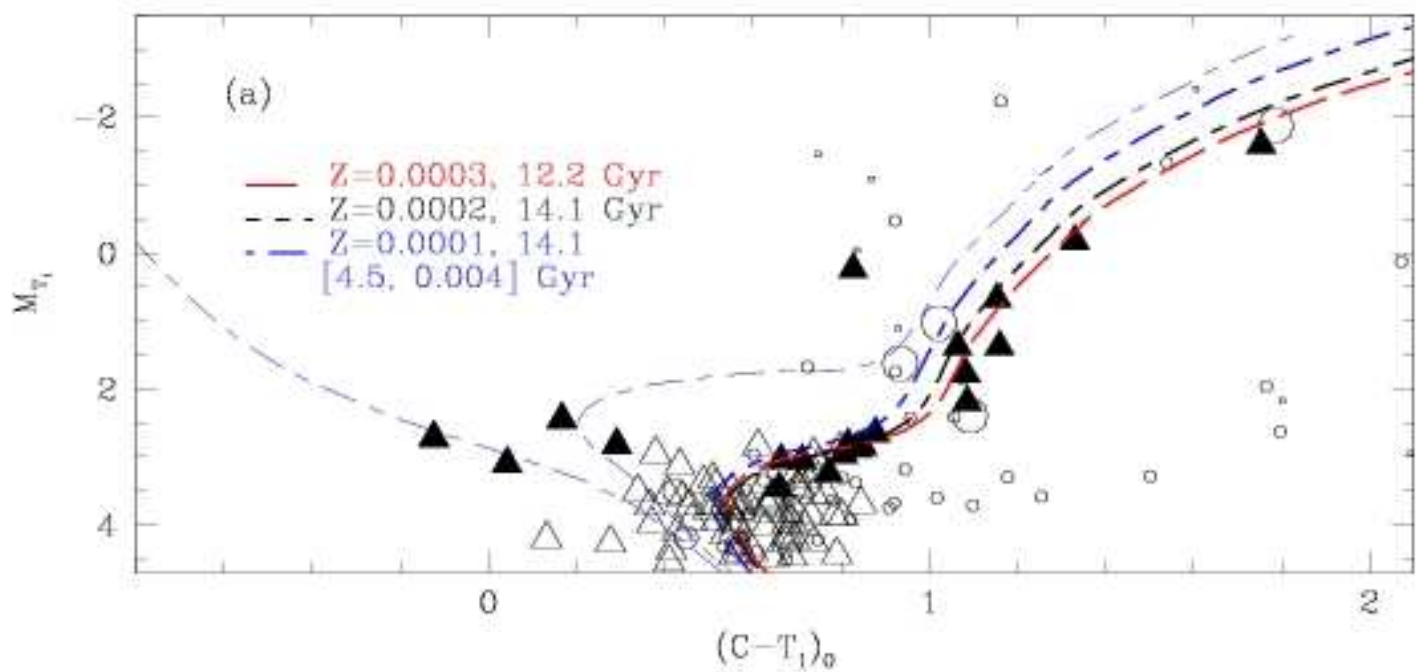
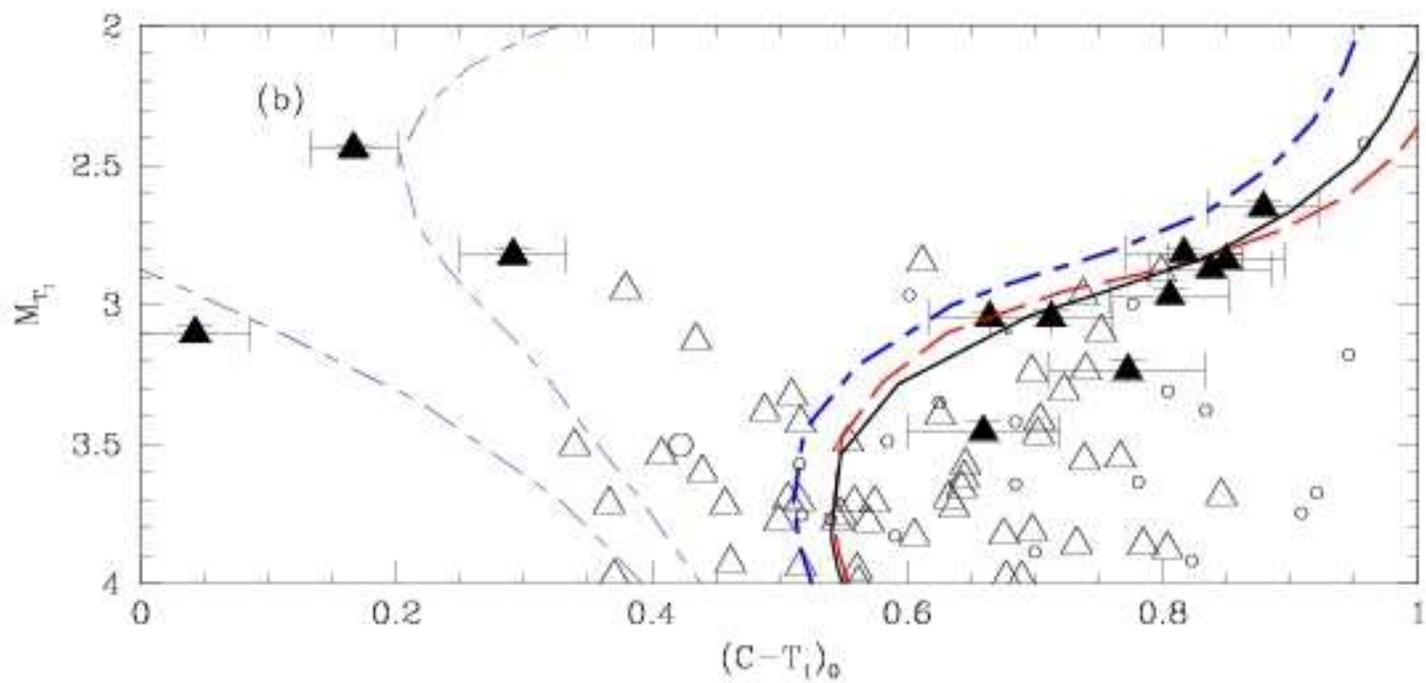
\end{document}